\documentstyle{europhys}

\def\And{{\rm and\ }}

\def\stars{\bigskip\centerline{***}\medskip}

\newif\ifboo \boofalse

\def\Review#1{\boofalse{\it #1},}
\def\Name#1{{\sc #1},}
\def\Vol#1{\ifboo Vol. {\bf #1}\else{\bf #1}\fi}
\def\Year#1{\ifboo #1\else(#1)\fi}
\def\Book#1{\bootrue{\it #1},}
\def\Page#1{\ifboo {\rm p. #1}\else{\rm #1}\fi}

\begin{document}
\euro{}{}{}{}
\Date{}
\shorttitle{P. \v{Z}upanovi\'{c}  {\em et al.} CRYSTAL STABILITY 
AND OPTICAL PROPERTIES ETC.}
\newcommand{\eq}[2]{\begin{equation} #1 \label{eq:#2} \end{equation}}
\newcommand{\req}[1]{(\ref{eq:#1})}
\title{Crystal stability and optical properties of 
organic chain compounds}

\author{P. \v{Z}upanovi\'{c}\inst{1}, A. Bjeli\v{s}\inst{2} \And 
 S. Bari\v{s}i\'{c}\inst{2}}
 \institute{ 
 \inst{1}  Department of Physics, Faculty of Science and Art,
University of Split\\
Teslina 10, 21000 Split, Croatia\\
\inst{2} Department of Physics, Faculty of Science, University        
 of Zagreb\\
POB 162, 10001 Zagreb, Croatia}
 \rec{}{} 
 \pacs{
 \Pacs{61}{50Lt}{ Crystal binding; cohesive energy }
 \Pacs{61}{66Hq}{Organic compounds}
 \Pacs{78}{30Jw}{Organic solids, polymers}
  }
\maketitle

\begin{abstract}

The solution to the long standing problem of the cohesion of 
organic chain compounds is proposed. We consider the tight-binding 
dielectric matrix with two electronic bands per chain, determine the 
corresponding hybridized collective modes, and show that three 
among them are considerably softened due to strong dipole-dipole and 
monopole-dipole interactions. By this we explain the unusual low frequency 
optical activity of TTF-TCNQ,  including the observed 10meV anomaly. 
The softening of the modes also explains the cohesion of the mixed-stack 
lattice, the fractional charge transfer almost independent of the material, 
and the formation of the charged sheets in some compounds.

\end{abstract}



\begin{figure}
\vspace*{7.5cm}
  \includegraphics{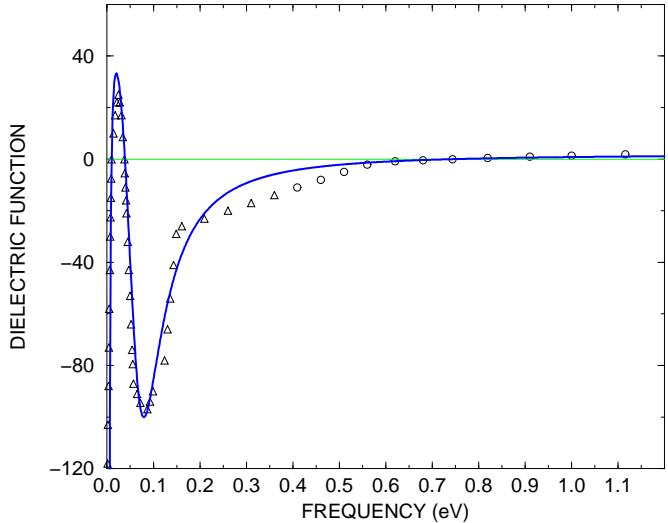}
\caption{The real part of the dielectric function for TTF-TCNQ at 100 K 
with the data taken from Refs. \protect\cite{jacobsen} (triangles)
and \protect\cite{tjgh} (circles). The full line is Re $\epsilon_M$ 
(eq. \protect\ref{epsM}) with the zeros at $\Omega_1$ = 10 meV, 
$\Omega_2$ = 750 meV, and $\Omega_3 = 6$ eV,  and poles at  
$\omega_{-t}$ = (57 + i42)meV and  $\omega_{+t} $=4.5 eV. The imaginary 
part in $\omega_{-t}$ is chosen to fit the experimental width of the pole in
Re $\epsilon_M$.}
\label{fig1}
\end{figure}

Ever since the discovery of the mixed stack organic chain compounds
(OCCs), some twenty five years ago, two important questions remained 
open \cite{bbrev}.  The first is the stability of their crystal 
structure, with a 
fractional charge transfer between the molecules of $n_a  \approx 0.6$ 
electrons per molecule. Furthermore, in some lattices
like TTF-TCNQ the equally 
charged (TTF or TCNQ) chains stack side by side, to form charged sheets, 
instead of alternating in both directions, as suggested by the simple 
Madelung argument. Such a simple alternation does occur in 
other lattices, like HMTTF-TCNQ, which, in spite of the 
radically different chain stacking, have similar $n_a$. 
 Neither the fractional charge 
transfer nor the formation of charged sheets are expected within
the standard ionic cohesion scheme which assumes the dominance of 
affinity-ionicity and electrostatic (Madelung) contributions. Besides,
calculations of cohesion energy within this scheme suggest that almost 
2 eV per donor-acceptor molecular pair is lacking. This "Madelung defect"
\cite{mb} clearly signifies that the cohesion at a scale of few eVs has to be
stabilized by some mechanism(s) beyond the standard ionic scheme. Although 
various suggestions \cite{mb}-\cite{zbb1} have been put forward, 
they have failed to explain all the aspects of the
cohesion, mostly because they could not account for the Madelung defect.
 On the other hand, Friedel \cite{friedel} pointed out that
the van der Waals interaction between the molecules
favors the formation of charged sheets observed in TTF-TCNQ. In this 
model the intramolecular (interband) and the intermolecular (intraband) 
charge fluctuations are separated. Such separation 
 however   fails in the low
frequency range in which TTF-TCNQ shows strong anomalies, observed in  
numerous microwave/infrared optical measurements 
\cite{gra}-\cite{bas} (fig.\ref{fig1}).

The origin of these anomalies is the second puzzle of OCCs. E. g. in 
TTF-TCNQ the frequency of the anomaly at 10 meV \cite{tjgh,jacobsen} is 
comparable to the value of the (pseudo)gap $\Delta_L$ produced by the 
$2k_F$ lattice instability at low temperatures ($T < 50 K$). However 
the anomalies in fig.\ref{fig1} can not be related to this instability, since 
they persist at high temperatures, where the $2k_F$ lattice fluctuations 
are practically absent. 

In this Letter we present a theory which for the first time 
 treats interband (van der Waals) and intraband correlations on equal
footing, and explains 
simultaneously the crystal stability and the optical anomalies of OCCs.
We investigate a simple model with two orbitals per donor and acceptor 
molecule, i. e. two bands per molecular chain, the lower bands being partially 
filled. Applying the recently proposed tight-binding formalism 
for the dielectric matrix in the random phase approximation (RPA) 
\cite{zbb2}, we determine the hybridized collective modes associated with the 
electron polarization processes, and the corresponding zero-point energy 
contributions to the cohesive energy. We show 
  that 
the above unusual cohesive and optical properties are related through the
presence of a soft interband collective mode and its hybridization with the 
intraband charge fluctuations.

In the  tight-binding formalism \cite{zbb2} the matrix elements of the screened 
Coulomb interaction $\bar{V}({\bf r},{\bf r'})$
are represented in terms of interband and intraband
polarization processes, and the corresponding system of RPA Dyson equations
reads
\begin{equation}
\label{a7}
\sum_{p_{g}}  \left[ 
 \delta_{p_{e}p_{g}}  - 
V_{p_{e}p_{g}}
( {\bf q})
\Pi_{p_{g}}({\bf q},\omega) \right] 
\bar{V}_{p_{g}{p}_{f}}( {\bf q},\omega)=
V_{p_{e}p_{f}}( {\bf q}).
\end{equation}
Here the matrix elements of the bare Coulomb interaction
$V({\bf r}-{\bf r'})$ are given by
\begin{equation}
V_{ p_e p_f    }( {\bf q})=
 \sum_{{\bf R}}e^{i{\bf q}({\bf R }+{\bf e}-{\bf f})}
 \int d{\bf r} \int d{\bf r'}
\varphi_{l_e}^*({\bf r}-{\bf R}-{\bf e})	 
\varphi_{n_f}^*({\bf r'}-{\bf f})V({\bf r}-{\bf r'})
\varphi_{l'_e}({\bf r}-{\bf R}-{\bf e})\varphi_{n'_f}({\bf r'}-{\bf f}).
\label{V}
\end{equation}
${\bf R}$ are lattice vectors, and ${\bf e}$ and ${\bf f}$ are
positions of the molecules within the unit cell. The latter take values
${\bf 0}$ and ${\bf a}/2$ for acceptor  and donor molecules respectively, 
with ${\bf R} = N_a{\bf a} + N_c{\bf c}$ for TTF-TCNQ, and 
${\bf R} = N_1({\bf a}/2 + {\bf c}) + N_2({\bf a}/2 - {\bf c})$
for HMTTF-TCNQ. Here $N_{a, c, 1, 2}$ are integers, and a simplified 
orthorombic symmetry with one donor and one acceptor chain per unit cell is 
assumed. The wave function $\varphi_{l_e}({\bf r} - {\bf R} - {\bf e})$ in 
eq.\ref{V} is $l_e$-th orbital at the molecular site ${\bf R} + {\bf e}$, 
and $p_e \equiv (l_e,l'_e)$ stays for the one-electron transition 
between  $l_e$ and $l'_e$ orbitals. We do not keep negligible 
contributions due to finite direct overlaps between orbitals from 
neighboring molecules. Having in mind the known electronic spectra
of e. g. TCNQ \cite{low} and TTF \cite{bh} molecules, we specify that
the orbitals which form lower (partially filled) and upper (empty) bands have 
the same, ${A_g}$ and ${B_{2u}}$, respective symmetries on both, 
acceptor and donor, sublattices. This means that the interband transitions 
on both families of chains are dipole active, with the dipole matrix elements 
$\mbox{\boldmath ${\mu}$}_{a(d)}$ oriented along the 
chain direction ${\bf b}$. With the standard assumption that all 
products $\varphi_{l_e}^*({\bf r})\varphi_{l_e^{'}}({\bf r})$ are real, 
we get  from  eq.\ref{a7} the dielectric matrix
\begin{equation}   
[\varepsilon]=
 \left[ 
\begin {array}{cccc}
1-V_{0_a0_a}\Pi_{0_a}&-V_{0_a0_d}\Pi_{0_d} & 
V_{0_a1_a}\Pi_{1_a} & -V_{0_a1_d}\Pi_{1_d}\\
-V_{0_d0_a}\Pi_{0_a}   &1- V_{0_d0_d}\Pi_{0_d} &  
       -V_{0_d1_a}\Pi_{1_a}       &-V_{0_d1_d}\Pi_{1_d}\\
-V_{1_a0_a}\Pi_{0_a}   &-V_{1_a0_d}\Pi_{0_d}&  
1-V_{1_a1_a}\Pi_{1_a}     &-V_{1_a1_d}\Pi_{1_d}\\
-V_{1_d0_a}\Pi_{0_a}   &-V_{1_d0_d}\Pi_{0_d}& 
-V_{1_d1_a}\Pi_{1_a}     &1-V_{1_d1_d}\Pi_{1_d}
\end{array}
\right]
\label{dm}
\end{equation}
for the microscopic response to the longitudinal electric field. 
Here indices $0_{a(d)}$ and $1_{a(d)}$ represent intraband and interband 
one-electron transitions on the acceptor (donor) chains, respectively.
$\Pi_{0_{a(d)}}$ and $\Pi_{1_{a(d)}}$ are intraband and interband bubble 
polarization diagrams, and $V's$ are the corresponding
matrix elements of eq. \ref{V}.
 Our main aim is to 
determine collective modes, i. e.  zeros of the
microscopic dielectric function $\varepsilon(\omega) = det[\varepsilon]$,
in the long wavelength limit ${\bf q} \rightarrow 0$. In this limit the  
intraband polarization diagrams are given by $\Pi_{0_{a(d)}}({\bf q}, \omega)
\approx  ({\bf qb})^2 \omega_{0a(d)}^2/4\pi  e^2 b^2 \omega^2$, while the 
interband polarization diagrams reduce to $\Pi_{1_{a(d)}}({\bf q}, \omega) 
\approx 2n_{a(d)}E_{a(d)}/(\omega^2 - E_{a(d)}^2)$, after taking 
into account that the energy gaps on both types of chains, 
$E_{a(d)}$ ($\approx 3 eVs$) \cite{pen}, 
are  considerably higher than the bandwidths ($\approx 0.5 eVs$) \cite{eps}.
Here $n_d = 2 - n_a$,
and $\omega_{0a(d)}$ and $E_{a(d)}$ are respectively plasmon frequencies
and differences between upper and lower electron band for the acceptor 
(donor) sublattices.  
Furthermore, the multipole expansion can be
performed for all Coulomb matrix elements except for the first neighbor
short-range interaction $V_{sr}$  in the ${\bf a}$ direction. In the 
${\bf q} \rightarrow 0$ limit
we keep for three types of the Coulomb matrix elements present in eq.\ref{dm},
namely those with intraband-intraband, intraband-interband and 
interband-interband scatterings, the leading, i. e. monopole-monopole 
($V_{0_e0_f} = 4\pi e^2/v_0 q^2$), monopole-dipole
($V_{0_e1_f} = 4\pi i e \mbox{\boldmath ${\mu}$}_f  \cdot 
{\bf q}/v_0 q^2$) and dipole-dipole ($V_{1_e1_f} = 4 \pi 
[3(\mbox{\boldmath ${\mu}$}_e  \cdot {\bf q})(\mbox{\boldmath ${\mu}$}_f  
\cdot {\bf q}) -
 q^2\mbox{\boldmath ${\mu}$}_e \cdot\mbox{\boldmath ${\mu}$}_f ]/3q^2
  + V_{sr}$), contributions respectively. $v_0$ is the
volume of the unit cell. Note that the
difference between the longitudinal (${\bf q}\| {\bf b}$) and transverse 
(${\bf q} \bot {\bf b}$) dipole-dipole matrix elements, which will appear in 
the further analysis, is independent of ${\bf q}$, as well as of the 
details of the crystal structure \cite{ck},
$V_{1_e1_f,l}-V_{1_e1_f,t}= (4 \pi \mu_e \mu_f)/v_0$.

From the form of the matrix (\ref{dm}), one may already recognize the 
hybridizations which lead to collective modes. We start with the
upper diagonal  $2\times 2$  block which represents intraband (metallic) 
polarizations with long range monopole-monopole interactions.
They give rise to two plasmon modes,  the hybrids of plasmons from
acceptor and donor sublattices. For strictly one-dimensional electron
bands, one of these hybridized modes is acoustic, while the other is optic,
with a frequency which vanishes in the limit ${\bf q} \rightarrow 0$ 
for ${\bf q} \bot {\bf b}$, and is equal to 
$\omega_{0} = \sqrt{\omega_{0a}^2 + \omega_{0d}^2}$
for ${\bf q} \| {\bf b}$ \cite{zbb1}. 

The lower diagonal $2\times 2$  block includes interband
processes, which, due to finite dipole matrix elements
at both types of molecules, induce long-range 
dipole-dipole interactions. The corresponding collective modes
\begin{equation}
 \omega_{\pm}^{2}=\frac{1}{2} 
\left[ \omega_{a}^2+\omega_{d}^{2} \pm 
\sqrt{\left(\omega_{a}^2-\omega_{d}^{2} \right)^2
 +16n_a n_d E_{a} E_{d} V_{1_{a}1_{d}}^2} \right]
\label{hdm}
\end{equation}
are hybrids of dipolar modes from two sublattices,
$\omega_{a(d)}^{2}= E_{a(d)} \left[ E_{a(d)}+ 
2 n_{a(d)} V_{1_{a(d)}1_{a(d)}} \right]$. 
The frequencies of the dipolar modes 
decrease from $\omega_{i,l}$ to  $\omega_{i,t}$ [$i \equiv a,d,\pm$]
as the orientation of ${\bf q}$ changes from ${\bf q}\| {\bf b}$
to ${\bf q} \bot {\bf b}$ \cite{zbb2} [and do not vary as  ${\bf q}$ 
rotates within the plane $({\bf a}, {\bf c})$]. Note that the frequency
differences $\omega^2_{i,l} - \omega^2_{i,t}$ are, like the corresponding
differences of dipole-dipole matrix elements, insensitive to the 
details of the crystal structure.

A further hybridization between plasmon $\omega_0$ and longitudinal
dipolar (\ref{hdm}) modes takes place through the off-diagonal 
$2\times 2$ blocks in the matrix (\ref{dm}). Due to the diverging long-range 
monopole-dipole Coulomb matrix elements, the latter do not vanish,
provided the lower bands are metallic. The frequencies of the three new 
hybridized longitudinal modes are solutions of the equation 
\begin{equation}
\omega^2(\omega^2 -\omega_{-l}^2)(\omega^2-\omega_{+l}^2) -
\omega_0^2(\omega^2 -\omega_{-t}^2)(\omega^2-\omega_{+t}^2) = 0,
\label{hpd}
\end{equation}
which we denote by $\Omega_{1,2,3}$. 
 Although the transverse dipolar frequencies 
$\omega_{\pm t}$ are not affected by the monopole-dipole interactions,
they enter into eq.\ref{hpd}, in agreement with our general results for
multiband metallic systems \cite{zbb2}. This has an important implication 
when $\omega_{-t}$, the frequency of the lower transverse mode,  is small, 
i. e. when 
\begin{equation}
V_{1_a1_{d},t} \approx \frac{\omega_{at}\omega_{dt}}
{2 \sqrt{n_a n_d E_a E_d}},
\label{cond}
\end{equation} 
which is, as we argue below, the case in TTF-TCNQ.
Then $\Omega_1$, the frequency of the lowest longitudinal mode from 
eq. \ref{hpd}, is also small and lies below $\omega_{-t}$,  i. e. 
\begin{equation}
\frac{\Omega_1^2}{\omega_{-t}^2}  \approx 
\left[1 + \left(\frac{\omega_{-l}\omega_{+l}}
{\omega_0 \omega_{+t}}\right)^2\right]^{-1}.
\label{ins}
\end{equation}
 The frequencies $\omega_{a(d)t}$ are presumably
comparable with the corresponding values in the homomolecular crystals of TCNQ 
and TTF molecules. Dielectric data for  TCNQ crystals give 
$\omega_{at} \approx E_a $ \cite{pen}. Analogous data for TTF are not 
available, but,  having in mind that the spectra of the 
TTF and TCNQ molecules are similar \cite{bh},   \cite{low}, and that both homomolecular crystals have 
similar cohesion energies  of $ 1 eV$ \cite{cruif}, we expect that 
$\omega_{dt} \approx E_d$, too. Thus the condition 
(\ref{cond}) is physically equivalent to the requirement
$E_a, E_d, V_{1_a1_{d}} \gg V_{1_a1_{a}},V_{1_d1_{d}}$, 
which in particular means that the dipolar coupling between donor and acceptor  
sublattices is much stronger than that within each sublattice. 
 Since this is not a property of the dipolar long-range 
lattice sums, i. e. the product of the long range parts of  $V_{1a1a}$ and
$V_{1d1d}$ is of the same  order of magnitude as the square of the long range 
part of  $V_{1a1d}$, the dominance of $V_{1a1d}$ is to be 
sought in the details of the short-range contributions from nearest neighboring 
pairs of chains.
We emphasize that, while the soft 
transverse mode $\omega_{-t}$ appears due to the strong, mainly local, 
dipolar fluctuations, the accompanying soft longitudinal mode $\Omega_1$ from 
eq.\ref{ins} appears due to their coupling of long-range intraband 
fluctuations.

The macroscopic dielectric function which corresponds to the matrix (\ref{dm})
is given by \cite{zbb2}
\begin{equation}
\epsilon_M = \frac{(\omega^2 - \Omega_1^2)(\omega^2 - \Omega_2^2)
(\omega^2 - \Omega_3^2)}
{(\omega^2 - \omega_{-t}^2)(\omega^2 - \omega_{+t}^2)}
\label{epsM}
\end{equation}
for ${\bf q} \parallel {\bf b}$, and $\epsilon_M = 1$ for 
${\bf q} \perp {\bf b}$. 
Since these functions correspond to the optical (transverse) dielectric 
functions for the orientation of the electric field parallel and perpendicular 
to the chain direction respectively, we can make a link with experiments,
keeping in mind that above expressions for $\epsilon_M$ do not contain 
contributions from interchain electron hoppings, molecular and lattice 
vibrations (including the $2k_F$ CDW collective modes), relaxation processes, 
etc. In particular, we assign the modes $\Omega_1$ and $\omega_{-t}$
to the respective excitations at  $\sim 10$ meV and $\sim 57$ meV, 
observed at 100 K in the infra-red measurements \cite{tjgh,jacobsen}, 
and the mode $\Omega_2$ to the excitation at $\sim 0.75$ eV, originally 
interpreted as a Drude edge \cite{gra,bas} (fig. \ref{fig1}). The modes
$\omega_{+t}$ and $\Omega_3$ are most likely situated in the frequency
range of few eVs, not investigated experimentally in detail
and usually attributed to the range of interband transitions. 
It is noteworthy to mention in this respect a strong anomaly at 
$\sim 4.5$ eV observed in the early reflectivity data \cite{gra}. 

The two most interesting modes, denoted here by $\Omega_1$ and
$\omega_{-t}$, invited various theoretical explanations. Here we list 
arguments in favor of the present one. First, our model predicts the optical 
activity in the chain ${\bf b}$ direction, in agreement with experiments 
\cite{tjgh,jacobsen}, and in contrast to the interpretation in terms of 
lattice vibrations \cite{gr}. The latter
are optically active only when polarized along the transverse ${\bf a}$ 
direction, in which the oppositely charged chains are aligned. Our suggestion 
that the regime (\ref{cond}) is appropriate for TTF-TCNQ is based on the 
observation that the frequencies $\Omega_1$, $\omega_{-t}$,  and $\Omega_{2}$ 
are much lower than those of other modes from eq.\ref{epsM}, which lie in the 
range of bare interband dipolar modes ($\sim 3$ eV). Furthermore, the 
experimentally determined difference of $\sim0.7$ eV between frequencies 
$\Omega_2$ and $\omega_{-t}$ \cite{tjgh}, which are the longitudinal and 
transverse edges of one hybridized branch (out of three), is comparable to 
the difference of $ \sim 1.5$ eV between analogous frequencies in, e.g., 
the TCNQ crystal \cite{pen} (although the characteristic values of latter 
are of the order of a few $eV$). Since these 
differences  measure the differences between longitudinal and transverse 
lattice sums in the matrix elements $V_{11}$, it follows
that the long-range contribution in $V_{1_a1_{d}}$ is comparable to those in 
$V_{1_a1_{a}}$ and $V_{1_d1_{d}}$, in accordance with the already mentioned
general result \cite{zbb2}. In other words, the small 
values of both $\omega_{-t}$ and $\Omega_2$ may originate only from the 
short-range ${\bf q}$-independent part of $V_{1_a1_{d}}$,
in agreement with our arguments below eq.\ref{ins}.

The onset of CDW order below $\sim 50 K$, which causes
the appearance of the well-known features in the far infra-red range
due to the CDW dynamics \cite{tcj}, have also important effects
on the modes of eq.\ref{epsM}. At first, the low-lying longitudinal 
mode $\Omega_1$ is eliminated \cite{tcj,bas} due to the formation of the 
insulating gap $\Delta_L$ at the Fermi level,
in agreement with our general predictions 
\cite{zbb2}. The modulation of the crystal structure in the insulating phase
below the temperature of the Peierls phase transition also causes a weak 
splitting of remaining transverse and longitudinal modes \cite{zbb2}, in 
analogy with Davydov splitting of molecular excitons \cite{dav}. Such splitting 
is indeed observed for the mode  $\omega_{-t}$ at $ \sim 50$ meV \cite{bas}. We
expect that more refined measurements would show the same splitting  for other
modes, e.g for  the mode $\Omega_2$ at $0.75 eVs$.

It remains to consider the contribution of the polarization modes 
(\ref{hpd}) to the cohesion. Extending the standard  RPA procedure
to the present multi-band system, one arrives at the cohesion energy
per unit cell
\begin{equation}
E_{coh} = E_{I} n_{a} - E_{M} n_{a}^2 + \frac{1}{2N}\sum_{{\bf q},j}
\omega_{j}({\bf q}) - E_{a} - E_{d},
\label{cohe}
\end{equation}
where $\omega_{j}$ are hybridized modes from eqs.\ref{hdm},\ref{hpd}. Due to 
the condition(\ref{cond}), the expression (\ref{cohe}) reduces for 
$n_a \leq 1$ to
\begin{equation}
E_{coh} \approx -(2C -  D - E_I)n_a + (C - \frac{A}{4} - 
\frac{D}{4} -  E_M)n_a^2 - D,
\label{coh}
\end{equation}
 Here $A (D) \equiv \sum_{\bf q} 
V^2_{1_{a(d)}1_{a(d)}}/2NE_{a(d)}$,
$C \equiv \sum_{\bf q} V^2_{1_{a}1_{d}}/N(E_{a}+E_{d})$,
and $E_I$ and $E_M$ are affinity-ionicity and Madelung energy respectively.
Noting that, in accordance with foregoing arguments, $A, D \ll C$, it follows
from eq.\ref{coh} that only with $2C - E_I > 0$ and $C - E_M > 0$ is partial 
charge transfer possible. In the case of TTF-TCNQ this means that, roughly, 
$C > 2 $eV, which just compensates the "Madelung defect" on one side, and is 
consistent with the condition for the $\omega_{-t}$ mode softening  
(\ref{cond}) on the other. As was already pointed out, the dominant 
contribution to $C$ is expected to come from the nearest-neighbor interaction 
$V_{sr}$ between acceptor and donor pairs, i. e. from the ${\bf a}$  direction 
which is indeed the direction of smallest compressibility in TTF-TCNQ 
\cite{deb}. This interchain coupling is expected to be
responsible for the fractional charge transfer in other OCCs as well. The 
fractional charge transfer is thus predicted to be relatively insensitive to 
the type of stacking in the ${\bf c}$ direction, which explains the similarity 
of the observed charge transfers in e. g. TTF-TCNQ and HMTTF-TCNQ. 

Let us now turn our attention to the relative stability of TTF-TCNQ and 
HMTTF-TCNQ lattices. Note that the short-range contribution from nearest 
neighbors in the ${\bf a}$ direction to the sum $C$ does not enter into the 
difference of corresponding cohesion energies (\ref{coh}), while the
contributions to the sums $A, D, C$ from the neighboring molecular pairs in 
the ${\bf c}$-direction which now become relevant, can be represented in the 
Friedel dipole-dipole form. Taking for simplicity $n_a = 1$, we come to the 
condition that TTF-TCNQ lattice is more stable than the HMTTF-TCNQ lattice
provided that  
\begin{equation}
\left[\left(\frac{\mu_a^2}{\sqrt{E_a}} - \frac{\mu_d^2}{\sqrt{E_d}}\right)^2 
+ 2 \frac{\mu_a^2 \mu_d^2 (\sqrt{E_a} - 
\sqrt{E_d})^2}{\sqrt{E_a E_d}(E_a + E_d)}\right]
\left(\frac{1}{c^6} - \frac{1}{[(a/2)^2 + c^2]^3}\right)
\geq E_M(H) - E_M(T).
\label{fri}
\end{equation}
 $E_M(H)$ and $ E_M(T)$ are Madelung energies of HMTTF-TCNQ and 
TTF-TCNQ lattices respectively. The left-hand side of 
this expression amounts to the second order 
van der Waals contribution. It is a slight generalization of the
original Friedel criterion, since  $E_a \neq E_d$  and the contribution 
from the diagonal ${\bf a}/2 + {\bf c}$ direction is taken into account.
 We note that the characteristic energies which are
responsible for the relative stability of   TTF-TCNQ and HMTTF-TCNQ lattices
are much smaller than those which determine the total cohesion, including the 
contributions from the coupling in the ${\bf a}$ direction which is decisive for
the fractional charge transfer. Indeed, the simple 
calculations of the differences of the Madelung \cite{zbb1} and the 
van der Walls energies in the point charge approximation by using the
numerical  values for  the molecular polarizabilities \cite{klymenko,ratner},
suggest that both differences are of the order of $0.1 eV$ per unit cell.

In summary, we have shown that the interband collective modes are responsible
for the crystal stability of OCCs, while the hybridization of of these modes 
with intraband plasmons explains the low frequency optical data in TTF-TCNQ
in particular. We argue, and this is the crucial point of our 
model, that the frequencies of three collective modes are rather low, i.e. 
below 1 eV. It is worthy to note that, since one of them falls into the 
frequency range of the CDW instabilities ($\leq 10$ meV), we are confronted 
with the interesting possibility of a mixing of these two types of correlation 
at low temperatures. This question is left for future investigations. Clearly, 
the concepts developed and successfully used here can also be applied to the 
other metallic systems with large molecular polarizabilities. 

\medskip\noindent
{\bf Acknowledgments}: We acknowledge the valuable remarks of J. Friedel.

\stars 
\vskip-12pt

\end{document}